\documentstyle[12pt]{article}

\begin{document}
\title{Masses of W and Z Bosons without Higgs}
\author{Bing An Li\\
Department of Physics and Astronomy, University of Kentucky\\
Lexington, KY 40506, USA}

\maketitle

\begin{abstract}
A Lagrangian of electroweak interactions without Higgs
is used to study
the contributions of quarks and leptons
to the masses of the W and
the Z bosons. It is shown that the $SU(2)\times U(1)$
symmetry is broken by both fermion masses and
axial-vector components of the
intermediate bosons. The masses of the W and
the Z bosons are obtained to be \(m^{2}_{W}={1\over2}g^{2}m^{2}_{t}\)
and \(m^{2}_{Z}=\rho m^{2}_{W}/cos^{2}\theta_{W}\) with
\(\rho\simeq 1\).
Two fixed gauge
fixing terms for W and Z boson fields are derived respectively. A coupling
between photon and Z boson is predicted. Massive neutrinos are required.
\end{abstract}

\newpage
The top quark has been discovered in Fermi laboratory[1], whose mass has been
determined to be
\begin{equation}
m_{t}=180\pm 12 GeV [2].
\end{equation}
The value of $m_{t}$ is at the same order of magnitude
as the masses of the W and the
Z bosons. As a matter of fact, before the discovery of the top quark
there were attempts of finding the relationship between top quark and
intermediate bosons by using various mechanism[3]. In this paper the
Lagrangian of GWS model of electroweak interactions without Higgs is
used to study the role of quarks and leptons in obtaining the masses
of the intermediate bosons. The Lagrangian of
electroweak (GWS model)
interactions without Higgs is
\begin{eqnarray}
\lefteqn{{\cal L}=
-{1\over4}A^{i}_{\mu\nu}A^{i\mu\nu}-{1\over4}B_{\mu\nu}B^{\mu\nu}
+\bar{q}\{i\gamma\cdot\partial-M\}q}
\nonumber \\
&&+\bar{q}_{L}\{{g\over2}\tau_{i}
\gamma\cdot A^{i}+g'{Y\over2}\gamma\cdot B\}
q_{L}+\bar{q}_{R}g'{Y\over2}\gamma\cdot Bq_{R}\nonumber \\
&&+\bar{l}\{i\gamma\cdot\partial-M_{f}\}l
+\bar{l}_{L}\{{g\over2}
\tau_{i}\gamma\cdot A^{i}-{g'\over2}\gamma\cdot B\}
l_{L}-\bar{l}_{R}g'\gamma\cdot B l_{R},
\end{eqnarray}
where $A^{i}_{\mu\nu}$ and $B_{\mu\nu}$ are electroweak boson fields,
M is the
mass matrix of the six quarks, $M_{l}$ is the lepton mass matrix,
$q_{L}$ is the left-handed quark
doublet, $q_{R}$ is the right handed quark field,
$l_{L}$ is the left-handed lepton
doublet, and $l_{R}$ is the right handed lepton field.
Summation over $q_{L}$, $q_{R}$, $l_{L}$, and $L_{R}$ is
implicated in Eq.(2).
This Lagrangian is used
to study the $SU(2)\times U(1)$ symmetry breaking
mechanism, the masses of the W and the Z bosons, and
the propagators of the intermediate boson fields in this paper.

We start from the doublet of t and b quarks. The Lagrangian
of this generation is
\begin{eqnarray}
\lefteqn{{\cal L}=
-{1\over4}A^{i}_{\mu\nu}A^{i\mu\nu}-{1\over4}B_{\mu\nu}B^{\mu\nu}
+\bar{t}\{i\gamma\cdot\partial-m_{t}\}t
+\bar{b}\{i\gamma\cdot\partial-m_{b}\}b}\nonumber \\
&&+\bar{\psi}_{L}\{{g\over2}\tau_{i}
\gamma\cdot A^{i}+g'{1\over6}\gamma\cdot B\}
\psi_{L}
+{2\over3}g'\bar{t}_{R}\gamma\cdot Bt_{R}
-{1\over3}g'\bar{b}_{R}\gamma
\cdot Bb_{R},
\end{eqnarray}
where
\[\psi_{L}=\left(\begin{array}{c}
                t\\b
                 \end{array} \right)_{L}.\]
The Lagrangian of the boson fields is obtained by integrating out
the fermion fields. For this purpose the quark part of
the Lagrangian(3) is rewritten as
\begin{equation}
{\cal L}=\bar{\psi}\{i\gamma\cdot\partial+\gamma\cdot v+
\gamma\cdot a\gamma_{5}-m\}\psi,
\end{equation}
where $\psi$ is the doublet of t and b quarks,
\[m=\left(\begin{array}{c}
          m_{t}\hspace{1cm}0\\
          0\hspace{1cm}m_{b}
          \end{array} \right),\]
\(v_{\mu}=\tau_{i}v^{i}_{\mu}+\omega_{\mu}\),
\(v^{1,2}_{\mu}={g\over4}A^{1,2}_{\mu}\),
\(v^{3}_{\mu}={g\over4}A^{3}_{\mu}+{g'\over4}B_{\mu}\),
\(\omega_{\mu}={g'\over6}B_{\mu}\),
\(a_{\mu}=\tau_{i}a^{i}_{\mu}\),
\(a^{1,2}_{\mu}=-{g\over4}
A^{1,2}_{\mu}\),
\(a^{3}_{\mu}=-{g\over4}A^{3}_{\mu}+{g'\over4}B_{\mu}\).
After finishing the integration over quark fields, the Lagrangian
of the intermediate boson fields is obtained in Euclidean space
\begin{equation}
{\cal L}_{E}=ln det{\cal D},
\end{equation}
where
\begin{equation}
{\cal D}=\gamma\cdot\partial-i\gamma\cdot v-i\gamma\cdot a\gamma_{5}+m.
\end{equation}
The real and imaginary parts of the Lagrangian(5) are
\begin{equation}
{\cal L}_{Re}={1\over2}ln det({\cal D}^{\dag}{\cal D}),\;\;\;
{\cal L}_{Im}={1\over2}ln det(\frac{{\cal D}}{{\cal D}^{\dag}}),
\end{equation}
where
\begin{equation}
{\cal D}^{\dag}=-
\gamma\cdot\partial+i\gamma\cdot v-i\gamma\cdot a\gamma_{5}+m.
\end{equation}
From Eqs.(6,7,8) it is learned that ${\cal L}_{Re}$ has even number
of $\gamma_{5}$ and normal parity. On the other hand, the number of
$\gamma_{5}$ in ${\cal L}_{Im}$ is odd and ${\cal L}_{Im}$ has
abnormal parity. Anomalous couplings between W, Z, and $\gamma$
should be revealed from ${\cal L}_{Im}$.

It is necessary to point out that
${\cal D}^{\dag}{\cal D}$ is a definite positive operator.
In terms of Schwinger's proper time method[4] ${\cal L}_{Re}$ is
expressed as
\begin{equation}
{\cal L}_{Re}={1\over2}\int d^{D}xTr\int^{\infty}_{0}{d\tau\over\tau}
e^{-\tau{\cal D}^{\dag}{\cal D}}.
\end{equation}
Inserting a complete set of plane waves and subtracting the divergence
at \(\tau=0\), we obtain
\begin{equation}
{\cal L}_{Re}={1\over2}\int d^{D}x
\frac{d^{D}p}{(2\pi)^{D}}Tr\int^{\infty}_{0}{d\tau\over \tau}
\{e^{-\tau{\cal D'}^{\dag}{\cal D}^{'}}-e^{-\tau\Delta_{0}}\},
\end{equation}
where
\begin{eqnarray}
\lefteqn{{\cal D'}=\gamma\cdot\partial+i\gamma\cdot p
-i\gamma\cdot v-i\gamma
\cdot a\gamma_{5}+m,}\nonumber \\
&&{\cal D}'^{\dag}=-\gamma\cdot\partial-i\gamma\cdot p+i\gamma\cdot v
-i\gamma\cdot a\gamma_{5}+m,\nonumber \\
&&{\cal D}'^{\dag}{\cal D}'=\Delta_{0}-\Delta,\nonumber \\
&&\Delta_{0}=p^{2}+m^{2}_{1},\;\;\;
m^{2}_{1}={1\over2}(m^{2}_{t}+m^{2}_{b}),\;\;\;
m^{2}_{2}={1\over2}(m^{2}_{t}-m^{2}_{b}),\nonumber \\
&&\Delta=\partial^{2}-(\gamma\cdot v-\gamma\cdot a\gamma_{5})
(\gamma\cdot v+\gamma\cdot a\gamma_{5})-i\gamma\cdot\partial
(\gamma\cdot v+\gamma\cdot a\gamma_{5})\nonumber \\
&&-i(\gamma\cdot v-\gamma\cdot a
\gamma_{5})\gamma\cdot\partial+2ip\cdot\partial
+2p\cdot(v+a\gamma_{5})
-i[\gamma\cdot v,m]\nonumber \\
&&+i\{\gamma\cdot a,m\}\gamma_{5}
-m^{2}_{2}\tau_{3}.
\end{eqnarray}
After the integration over $\tau$, ${\cal L}_{Re}$ is expressed as
\begin{equation}
{\cal L}_{Re}={1\over2}\int d^{D}x\frac{d^{D}p}{(2\pi)^{D}}
\sum^{\infty}_{n=1}{1\over n}\frac{1}{(p^{2}+m^{2}_{1})^{n}}
Tr\Delta^{n}.
\end{equation}
Due to the property of ${\cal L}_{Im}$ mentioned above,
${\cal L}_{Im}$ doesn't contribute to the masses of W and Z bosons
at least at the tree level of the boson fields. Therefore, the study
of ${\cal L}_{Im}$ is beyond the scope of this paper.

${\cal L}_{Re}$ is used to investigate
the symmetry breaking mechanism, the masses of the intermediate bosons,
and the propagators of the boson fields.
Obviously, the $SU(2)\times U(1)$ symmetry is broken by both
quark and lepton masses.
However, another symmetry breaking mechanism is needed.
{\bf Due to parity
nonconservation the intermediate bosons have both  vector and
axial-vector components which are written in the forms of
v and a in
Eq.(4). This fact makes the intermediate boson fields different
from photon and gluon fields which are vector fields.}
In this paper we investigate whether this property of the intermediate
boson fields results in another
$SU(2)\times U(1)$ symmetry breaking.
From the expression of $\Delta$(11)
it is seen that in company with fermion mass
the vector component v of the intermediate boson fields
appears in a commutator $[v,m]$, while the axial-vector
component a appears in an anticommutator
$\{a,m\}$. [v,m] provides the mass for
$W^{\pm}$ only, $\{a,m\}$ provides the masses for both W and Z bosons.
Indeed,
the axial-vector component of the intermediate boson
leads to a new
$SU(2)\times U(1)$ symmetry breaking mechanism.
The details of the symmetry breaking by both
fermion masses and
axial components of the intermediate boson fields are shown in the
calculations of $m_{W}$ and $m_{Z}$.

In terms of the Lagrangian(12) the masses of the
intermediate bosons are calculated.
The Lagrangian(12) shows that the electric $U(1)$ symmetry is
remained, therefore, there is massless boson field, which is
the photon field.
The terms related to the masses only
is separated from Eq.(12)
\begin{eqnarray}
\lefteqn{{\cal L}_{M}={1\over2}\int d^{D}x\int\frac{d^{D}p}{(2\pi)^{D}}
\sum^{\infty}_{n=1}{1\over n}\frac{1}{(p^{2}+m^{2}_{1})^{n}}Tr\{
-(\gamma\cdot v-\gamma\cdot a\gamma_{5})}\nonumber \\
&&(\gamma\cdot v+\gamma\cdot
a\gamma_{5})+2p\cdot(v+a\gamma_{5})
+i[m,\gamma\cdot v]+i\{m,
\gamma\cdot a\}\gamma_{5}-m^{2}_{2}\tau_{3}\}^{n}.
\end{eqnarray}
The contributions of the fermion masses to $m_{W}$ and $m_{Z}$
are needed to be calculated to all orders.
It is found that the series of the fermion masses are convergent to
analytic functions.
The mass terms of the intermediate bosons are obtained from Eq.(13).
In Minkowski space it is expressed as
\begin{eqnarray}
\lefteqn{{\cal L}_{M}={1\over2}\frac{N_{C}}{(4\pi)^{2}}\{{D\over4}
\Gamma(2-{D\over2})(4\pi)^{{\epsilon\over2}}({\mu^{2}\over m^{2}_{1}})
^{{\epsilon\over2}}+{1\over2}[1-ln(1-x)-
(1+{1\over x}){\sqrt{x}\over2}
ln\frac{1+\sqrt{x}}{1-\sqrt{x}}]\}m^{2}_{1}g^{2}
\sum^{2}_{i=1}A^{i}_{\mu}A^{i\mu}}\nonumber \\
&&+{1\over2}\frac{N_{C}}{(4\pi)^{2}}\{{D\over4}
\Gamma(2-{D\over2})(4\pi)^{{\epsilon\over2}}({\mu^{2}\over m^{2}_{1}})
^{{\epsilon\over2}}-{1\over2}[ln(1-x)+\sqrt{x}
ln\frac{1+\sqrt{x}}{1-\sqrt{x}}]\}m^{2}_{1}(g^{2}+g'^{2})
Z_{\mu}Z^{\mu},
\end{eqnarray}
where $N_{C}$ is the number of colors and
\(x=({m^{2}_{2}\over m^{2}_{1}})^{2}\). It is necessary to point
out that
\begin{equation}
a^{3}_{\mu}=\sqrt{g^{2}+g'^{2}}Z_{\mu}.
\end{equation}
In the same way, other two generations of quarks,
\(\left(\begin{array}{c}
         u\\d
        \end{array} \right) \)
and
\(\left(\begin{array}{c}
         c\\s
        \end{array} \right) \),
and three generations of leptons
\(\left(\begin{array}{c}
         \nu_{e}\\e
        \end{array} \right) \),
\(\left(\begin{array}{c}
         \nu_{\mu}\\\mu
        \end{array} \right) \), and
\(\left(\begin{array}{c}
         \nu_{\tau}\\\tau
        \end{array} \right) \)
contribute to the masses of W and Z bosons too. By changing the
definitions of $m^{2}_{1}$ and x to the quantities of other
generations in Eq.(14),
the contributions of the other two quark generations
are found. Taking off the factor $N_{C}$ and changing $m^{2}_{1}$ and
x to corresponding quantities of leptons, the contributions of the
leptons to $m_{W}$ and $m_{Z}$ are obtained.
In this paper the effects of CKM matrix are not taken into account.
The final expressions of the masses of W and Z bosons
are the sum of the contributions of the three quark and the three
lepton generations.
It is learned from the processes deriving Eq.(14) that
\begin{enumerate}
\item The field $sin\theta_{W}A^{3}_{\mu}+cos\theta_{W}B_{\mu}$(photon
field) is massless;
\item Both W and Z bosons gain masses from fermion masses;
\item In Eq.(13) there are four different
types of terms contributing to $m_{W}$ and $m_{Z}$. It is found that
$m_{Z}$ is resulted in $\{a_{\mu},m\}$ only.
Without $\{a_{\mu}, m\}$ Z boson is massless.
\end{enumerate}
The conclusion is that the fermion masses and
the axial components of the
intermediate boson fields cause the $SU(2)\times U(1)$ symmetry
breaking and makes both W and Z bosons massive.

Now the W and Z bosons are massive. It is necessary to study their
propagators to see whether they have right behavior for
renormalization at high energy.
Up to all orders of fermion masses,
the kinetic terms of the intermediate boson fields
are obtained from the Lagrangian(12). In Minkowski space it is
expressed as
\begin{eqnarray}
\lefteqn{{\cal L}_{K}=-{1\over4}\sum_{i=1,2}
(\partial_{\mu}A^{i}_{\nu}-
\partial_{\nu}A^{i}_{\mu})^{2}\{1+{1\over(4\pi)^{2}}g^{2}
\sum_{q,l}N[{D\over12}
\Gamma(2-{D\over2})(4\pi)^{{\epsilon\over2}}({\mu^{2}\over m^{2}_{1}})
^{{\epsilon\over2}}-{1\over6}+f_{1}]\}}\nonumber \\
&&-{1\over4}(\partial_{\mu}A^{3}_{\nu}-\partial_{\nu}A^{3}_{\mu})^{2}
\{1+\frac{1}{(4\pi)^{2}}g^{2}\sum_{q,l}N
[{D\over12}\Gamma(2-{D\over2})
(4\pi)^{{\epsilon\over2}}({\mu^{2}\over m^{2}_{1}})
^{{\epsilon\over2}}-{1\over6}+{1\over6}f_{2}]\}\nonumber \\
&&-{1\over4}(\partial_{\mu}B_{\nu}-\partial_{\nu}B_{\mu})^{2}
\{1+\frac{N_{C}}{(4\pi)^{2}}g'^{2}\sum_{q}[{11\over9}
{D\over12}\Gamma(2-{D\over2})
(4\pi)^{{\epsilon\over2}}({\mu^{2}\over m^{2}_{1}})
^{{\epsilon\over2}}-{1\over6}+{11\over54}f_{2}-{1\over18}f_{3}
]\}\nonumber \\
&&+\frac{1}{(4\pi)^{2}}g'^{2}\sum_{l}[
{D\over4}\Gamma(2-{D\over2})
(4\pi)^{{\epsilon\over2}}({\mu^{2}\over m^{2}_{1}})
^{{\epsilon\over2}}-{1\over6}+{1\over2}f_{2}+{1\over6}f_{3}
]\}\nonumber \\
&&-\frac{1}{(4\pi)^{2}}{gg'\over12}(\partial_{\mu}A^{3}_{\nu}
-\partial_{\nu}A^{3}_{\mu})(\partial^{\mu}B^{\nu}-\partial^{\nu}B^{\mu})
\{N_{G}-{2\over3}N_{C}\sum_{q}f_{3}+2\sum_{l}f_{3}\}\nonumber \\
&&+\frac{1}{(4\pi)^{2}}N_{G}
{g^{2}\over12}\sum_{i=1,2}(\partial^{\mu}A^{i}_{\mu})
^{2}+\frac{1}{(4\pi)^{2}}N_{G}
{1\over12}(g^{2}+g'^{2})(\partial_{\mu}
Z^{\mu})^{2},
\end{eqnarray}
where $\sum_{q}$ and $\sum_{l}$ stand for summations of
generations of quarks and leptons respectively,
\(N=N_{C}\) for q and \(N=1\) for l, \(N_{G}=3N_{C}+3\),
x depends on fermion generation and is defined in Eq.(14)
for one generation,
\begin{eqnarray}
\lefteqn{f_{1}={4\over9}-{1\over6x}
-{1\over6}ln(1-x)
+{1\over4\sqrt{x}}({1\over3x}-1)
ln\frac{1+\sqrt{x}}{1-\sqrt{x}},}\nonumber \\
&&f_{2}=-ln(1-x),\;\;\;
f_{3}=
{1\over2}{1\over\sqrt{x}}ln\frac{1+\sqrt{x}}
{1-\sqrt{x}}.
\end{eqnarray}
Following results are obtained from Eq.(16)
\begin{enumerate}
\item The boson fields and the coupling constants g and g'
have to be redefined by multiplicative renormalization
\begin{eqnarray}
A^{1,2}_{\mu}\rightarrow Z^{{1\over2}}_{1}A^{1,2}_{\mu},
A^{3}_{\mu}\rightarrow Z^{{1\over2}}_{3}A^{3}_{\mu},
B_{\mu}\rightarrow Z^{{1\over2}}_{B}B{\mu},\nonumber \\
g_{1}=Z^{-{1\over2}}_{1}g,\;\;\;g_{3}=Z^{-{1\over2}}_{3}g,\;\;\;
g_{B}=Z^{-{1\over2}}_{B}g',
\end{eqnarray}
where
\begin{eqnarray}
\lefteqn{Z_{1}=
1+{1\over(4\pi)^{2}}g^{2}\sum_{q,l}N
[{D\over12}
\Gamma(2-{D\over2})(4\pi)^{{\epsilon\over2}}
({\mu^{2}\over m^{2}_{1}})
^{{\epsilon\over2}}-{1\over6}+f_{1}]}\nonumber \\
&&Z_{3}=1+\frac{1}{(4\pi)^{2}}g^{2}\sum_{q,l}N[
{D\over12}\Gamma(2-{D\over2})
(4\pi)^{{\epsilon\over2}}({\mu^{2}\over m^{2}_{1}})
^{{\epsilon\over2}}-{1\over6}+{1\over6}f_{2}]\nonumber \\
&&Z_{B}=1+\frac{N_{C}}{(4\pi)^{2}}g'^{2}\sum_{q}[{11\over9}
{D\over12}\Gamma(2-{D\over2})
(4\pi)^{{\epsilon\over2}}({\mu^{2}\over m^{2}_{1}})
^{{\epsilon\over2}}-{1\over6}+{11\over54}f_{2}-{1\over18}f_{3}
]\nonumber \\
&&+\frac{1}{(4\pi)^{2}}g'^{2}\sum_{l}[
{D\over4}\Gamma(2-{D\over2})
(4\pi)^{{\epsilon\over2}}({\mu^{2}\over m^{2}_{1}})
^{{\epsilon\over2}}-{1\over6}+{1\over2}f_{2}+{1\over6}f_{3}
].
\end{eqnarray}
The divergent terms in $Z_{1}$ and $Z_{3}$ are the same.
\item There is a crossing term between
$A^{3}_{\mu}$ and $B_{\mu}$, which is written as
\begin{eqnarray}
\lefteqn{g_{3}g_{B}(\partial_{\mu}A^{3}_{\nu}-\partial_{\nu}A^{3}_{\mu})
(\partial_{\mu}B_{\nu}-\partial_{\nu}B_{\mu})
=e^{2}(\partial_{\mu}A_{\nu}-\partial_{\nu}A_{\mu})^{2}
-e^{2}(\partial_{\mu}Z_{\nu}-\partial_{\nu}Z_{\mu})^{2}}
\nonumber \\
&&+e(g_{3}cos\theta_{W}-g_{B}sin\theta_{W})
(\partial_{\mu}A_{\nu}-\partial_{\nu}A_{\mu})
(\partial_{\mu}Z_{\nu}-\partial_{\nu}Z_{\mu}),
\end{eqnarray}
where \(sin\theta_{W}=\frac{g_{B}}{\sqrt{g^{2}_{B}+g^{2}_{3}}}\),
\(cos\theta_{W}=\frac{g_{3}}{\sqrt{g^{2}_{B}+g^{2}_{3}}}\),
and \(e=\frac{g_{3}g_{B}}{\sqrt{g^{2}_{B}+g^{2}_{3}}}\).
Therefore, the photon and the Z fields are needed to be renormalized
again
\begin{equation}
(1+{\alpha\over4\pi}f_{4})^{{1\over2}}A_{\mu}\rightarrow A_{\mu},\;\;\;
(1-{\alpha\over4\pi}f_{4})^{{1\over2}}Z_{\mu}\rightarrow Z_{\mu},
\end{equation}
where
\[f_{4}={1\over3}N_{G}-{2\over3}\sum_{q}f_{3}+{2\over3}\sum_{l}f_{3}.\]
After these renormalizations(18,21),
${\cal L}_{K}$(16) is rewritten as
\begin{eqnarray}
\lefteqn{{\cal L}_{K}=
-{1\over 4}(\partial_{\mu}A_{\nu}-\partial_{\nu}
A_{\mu})^{2}
-{1\over 4}\sum_{i=1,2}(\partial_{\mu}A^{i}_{\nu}-\partial_{\nu}
A^{i}_{\mu})^{2}
-{1\over 4}(\partial_{\mu}Z_{\nu}-\partial_{\nu}
Z_{\mu})^{2}}\nonumber \\
&&-{1\over4}\frac{\alpha}{4\pi}
({g_{3}\over g_{B}}-{g_{B}\over g_{3}})
(1-\frac{\alpha^{2}}{(4\pi)
^{2}}f^{2}_{4})^{-{1\over2}}f_{4}
(\partial_{\mu}A_{\nu}-
\partial_{\nu}A_{\mu})(\partial^{\mu}Z^{\nu}-\partial^{\nu}Z^{\mu})
\nonumber \\
&&-\frac{N_{G}}{(4\pi)^{2}}
{g^{2}\over12}(\partial^{\mu}A^{i}_{\mu})
^{2}-\frac{N_{G}}{(4\pi)^{2}}(1-{\alpha\over4\pi}f_{4})^{-1}
{1\over12}(g^{2}_{3}+g^{2}_{B})(\partial_{\mu}
Z^{\mu})^{2}.
\end{eqnarray}
\item
The interaction between photon and Z boson is predicted in Eq.(22).
For very small neutrino masses it is derived from Eq.(17)
\begin{equation}
f_{3}=-ln({m_{\nu}\over m_{l}}).
\end{equation}
Therefore, if neutrino is massless
$f_{4}$ is logarithmic divergent. This divergence is in contradiction
with that the physical coupling between photon and z boson must be
finite.
Therefore, this theory requires massive neutrinos.
\item Fixed gauge fixing terms for W- and Z- fields are generated(22)
\begin{eqnarray}
-{1\over4}\xi_{W}(\partial^{\mu}A^{i}_{\mu})^{2},\;\;\;
-{1\over4}\xi_{Z}(\partial^{\mu}Z_{\mu})^{2},\nonumber \\
\xi_{W}=\frac{N_{G}}{(4\pi)^{2}}{g^{2}_{1}\over3},\;\;\;
\xi_{Z}=\frac{N_{G}}{(4\pi)^{2}}(1-{\alpha\over4\pi}f_{4})^{-1}
{1\over3}(g^{2}_{3}+g^{2}_{B}).
\end{eqnarray}
The propagator of W field is derived from Eq.(22)
\begin{equation}
\frac{i}{q^{2}-m^{2}_{W}}\{-g_{\mu\nu}+\frac{q_{\mu}q_{\nu}}{q^{2}}
\}-\frac{i}{\xi_{W}q^{2}-m^{2}_{W}}\frac{q_{\mu}q_{\nu}}{q^{2}}.
\end{equation}
Changing the index W to Z in Eq.(25),
the propagator of Z boson field
is obtained. Obviously, due to the gauge fixing terms the propagators
of W an Z bosons do not affect the renormalizability of the theory(2).
It is necessary to point out that the W and the Z fields are massive
and no longer gauge fields.
The "gauge fixing"
terms of W and Z bosons(24) are derived from this theory and they
are not obtained by choosing gauge. They
are really fixed.
\end{enumerate}

Now we can study the values of the masses of W and Z bosons.
After the renormalizations(18,21) there are still divergences in the
mass formulas of $m_{W}$ and $m_{Z}$(14). The boson fields are
already renormalized and the kinetic terms of the boson fields
are already in the standard form. Therefore, the divergences
in the formulas of $m_{W}$ and $m_{Z}$
cannot be absorbed by the boson fields. On the other hand,
the divergences in Eq.(14) are fermion mass dependent, while
the coupling constants should be the same for all fermion
generations. It is difficult that these divergences are absorbed
by the coupling constants. In Eq.(14) the fermion masses are bare
physical quantities. It is reasonable to redefine the fermion
masses by multiplicative renormalization
\begin{eqnarray}
Z_{m}m^{2}_{1}=m^{2}_{1,P},\nonumber \\
Z_{m}=\frac{N}{(4\pi)^{2}}\{N_{G}{D\over4}\Gamma(2-{D\over2})
(4\pi)^{{\epsilon\over2}}({\mu^{2}\over m^{2}_{1}})
^{{\epsilon\over2}}+{1\over2}[1-ln(1-x)-
(1+{1\over x}){\sqrt{x}\over2}
ln\frac{1+\sqrt{x}}{1-\sqrt{x}}]\},
\end{eqnarray}
for each generation of fermions. The index "P" is omitted in the rest
of the paper.
Now the mass of W boson is obtained from Eq.(14)
\begin{equation}
m^{2}_{W}={1\over2}g^{2}\{m^{2}_{t}+m^{2}_{b}+m^{2}_{c}+m^{2}_{s}
+m^{2}_{u}+m^{2}_{d}+m^{2}_{\nu_{e}}+m^{2}_{e}
+m^{2}_{\nu_{\mu}}+m^{2}_{\mu}+m^{2}_{\nu_{\tau}}+m^{2}_{\tau}\}
\end{equation}
Obviously, the top quark mass dominates the $m_{W}$
\begin{equation}
m_{W}={g\over\sqrt{2}}m_{t}.
\end{equation}
Using the values \(g=0.642\) and \(m_{t}=180\pm12 GeV\)[2], it is found
\begin{equation}
m_{W}=81.71(1\pm0.067) GeV,
\end{equation}
which is in excellent agreement with data $80.33\pm0.15$GeV[2].

Using Eqs.(14,18,21),
the mass formula of the Z boson
is written as
\begin{equation}
m^{2}_{Z}=\rho m^{2}_{W}(1+{g^{2}_{B}\over g^{2}_{1}}),
\end{equation}
where
\begin{eqnarray}
\lefteqn{\rho=(1-{\alpha\over4\pi}f_{4})^{-1}\sum_{q,l}N
\{{D\over4}
\Gamma(2-{D\over2})(4\pi)
^{{\epsilon\over2}}({\mu^{2}\over m^{2}_{1}})
^{{\epsilon\over2}}-{1\over2}[ln(1-x)+\sqrt{x}
ln\frac{1+\sqrt{x}}{1-\sqrt{x}}]\}}
\nonumber \\
&&/\sum_{q,l}N\{{D\over4}
\Gamma(2-{D\over2})(4\pi)^{{\epsilon\over2}}({\mu^{2}\over m^{2}_{1}})
^{{\epsilon\over2}}+{1\over2}[1-ln(1-x)-
(1+{1\over x}){\sqrt{x}\over2}
ln\frac{1+\sqrt{x}}{1-\sqrt{x}}]\}
\frac{g^{2}_{3}+g^{2}_{B}}
{g^{2}_{1}+g^{2}_{B}}.
\end{eqnarray}
Comparing
to the infinites in Eqs.(18,19,31), the finite terms can be
ignored in Eqs.(18,19,31). We have
\begin{eqnarray}
g_{1}=g_{3}\equiv g_{A},\nonumber \\
m^{2}_{Z}=\rho m^{2}_{W}/cos^{2}\theta_{W},\;\;\;
cos\theta_{W}=g_{A}/\sqrt{g^{2}_{A}+g^{2}_{B}},\;\;\;
\rho=(1-{\alpha\over4\pi}f_{4})^{-1},
\end{eqnarray}
where $g_{A}$ and $g_{B}$ are g and g' of the GWS model respectively.
The finiteness of the $\rho$ factor requires a finite $f_{4}$.
Once again massive neutrinos are required.
Due to the smallness of the factor ${\alpha\over4\pi}$
in the reasonable ranges of the quark masses
and the upper
limits of neutrino masses we expect
\begin{equation}
\rho\simeq 1.
\end{equation}
Therefore,
\begin{equation}
m_{Z}=m_{W}/cos\theta_{W}
\end{equation}
is a good approximation.

Introduction of a cut-off leads to
\begin{equation}
{D\over 4}
\Gamma(2-{D\over2})(4\pi)^{{\epsilon\over2}}({\mu^{2}\over m^{2}_{1}})
^{{\epsilon\over2}}\rightarrow
ln(1+{\Lambda^{2}\over
m^{2}_{1}})-1+\frac{1}{1+{\Lambda^{2}\over m^{2}_{1}}}.
\end{equation}
Taking $\Lambda\rightarrow\infty$,
Eq.(32) is obtained.
On the other hand,
the cut-off might be estimated by the value of the $\rho$
factor. The cut-off can be
considered as
the energy scale of unified electroweak theory.

To conclude, a Lagrangian without Higgs is investigated.
The $SU(2)\times U(1)$ symmetry is broken by both the
fermion masses and the axial-vector components of the intermediate
boson fields. W and Z bosons gain masses. Two fixed gauging fixing
terms for W- and Z- fields are derived, which make the propagators
of W- and Z- fields have right behaviors for renormalization. A
coupling between photon and Z boson is predicted. The finiteness
of $m_{Z}$, the $\gamma-Z$ coupling, and the fixed gauging fixing
terms require massive neutrinos.
By renormalizing the fermion masses, a mass formula
of W boson is obtained and in excellent agreement with data. A
\(\rho\simeq1\) factor is predicted in the mass formula of Z boson.
The formalism developed in this paper can be used to study the
couplings between intermediate bosons($\gamma$, W, Z, and gluons).

This research was partially
supported by DOE Grant No. DE-91ER75661.

\end{document}